\documentclass[aps,twocolumn,floats,prd,nofootinbib,superscriptaddress,10pt]{revtex4-2}

\usepackage[utf8]{inputenc}
\usepackage{graphicx}
\usepackage{dcolumn}
\usepackage{amsmath}
\usepackage{amssymb}
\usepackage{amsfonts}
\usepackage{bm}
\usepackage{hyperref}
\usepackage{lipsum}
\usepackage{xcolor}
\usepackage{calc}
\usepackage{accents}
\usepackage{comment}
\usepackage{float}
\newcommand{\Odcdm}{\Omega_{\rm{dcdm}}^{\rm{ini}} }

\newcommand{\eps}{\varepsilon}
\newcommand\wdm{{\rm{wdm}}}

\newcommand\dcdm{{\rm{dcdm}}}

\newcommand{\Planck}{{\sc Planck}}
\newcommand{\BAO}{{\sc BAO}}
\newcommand{\FS}{{\sc FS}}
\newcommand{\Pantheon}{{\sc Pantheon}}

\begin{document}

\preprint{APS/123-QED}

\title{Implications of the $S_8$ tension for decaying dark matter with warm decay products}

\author{Guillermo Franco Abell\'an}
\email{guillermo.franco-abellan@umontpellier.fr }
\author{Riccardo Murgia}
\email{riccardo.murgia@umontpellier.fr}
\author{Vivian Poulin}
\email{vivian.poulin@umontpellier.fr }
\author{Julien Lavalle}
\email{lavalle@in2p3.fr }
\affiliation{
 Laboratoire Univers \& Particules de Montpellier (LUPM), Universit\'e de Montpellier (UMR-5299) \\ Place Eugène Bataillon, F-34095 Montpellier Cedex 05, France 
}

\date{\today}
\begin{abstract}

Recent weak lensing surveys have revealed that the direct measurement of the parameter combination $S_8\equiv\sigma_8(\Omega_m/0.3)^{0.5}$ --  where $\sigma_8$ is a  measure of the amplitude of matter fluctuations on 8 $h^{-1}$Mpc scales -- is $\sim3\sigma$ discrepant with the value reconstructed from cosmic microwave background (CMB) data assuming the $\Lambda$CDM model. In this article, we show that it is possible to resolve the tension if dark matter (DM) decays with a lifetime of $\Gamma^{-1} \simeq  55 \ \text{Gyrs}$ into one massless and one massive product, and transfers a fraction $\varepsilon\simeq 0.7 \ \%$ of its rest mass energy to the massless component. The velocity-kick received by the massive daughter leads to a suppression of gravitational clustering below its free-streaming length, thereby reducing the $\sigma_8$ value as compared to that inferred from the standard $\Lambda$CDM model, in a similar fashion to massive neutrino and standard warm DM.  Contrarily to the latter scenarios, the time-dependence of the power suppression and the free-streaming scale allows the 2-body decaying DM scenario to accommodate CMB, baryon acoustic oscillation, growth factor and un-calibrated supernova Ia data. We briefly discuss implications for DM model building, galactic small-scale structure problems and the recent Xenon-1T excess. Future experiments measuring the growth factor to high accuracy at $0\lesssim z\lesssim1$ can further test this scenario.
 
\end{abstract}

\maketitle

\label{sec_dcdm:Intro}
\section{Introduction} The standard $\Lambda$-cold dark matter ($\Lambda$CDM) cosmological model 
provides a remarkable fit to a wide variety of observables, such as big bang nucleosynthesis (BBN), the cosmic microwave background (CMB), large-scale structures (LSS), baryonic acoustic oscillations (BAO), and uncalibrated supernovae of type Ia (SNIa) (see e.g.~\cite{Scott:2018adl,Akrami:2018vks} for reviews).
Nevertheless, tremendous experimental developments have revealed curious discrepancies between different probes. At the heart of this study is the growing tension between the cosmological and local determination of the amplitude of the matter fluctuations on 8 $h^{-1}$/Mpc scales, typically described through the parameter combination $S_8\equiv\sigma_8(\Omega_m/0.3)^{0.5}$.
Within $\Lambda$CDM, the latest $S_8$ value inferred from a fit to CMB data~\cite{Cosmo_collaboration2018planck} is $\sim 2-3\sigma$ higher than that measured by a host of weak lensing surveys such as CFHTLenS~\cite{Heymans:2013fya}, HSC~\cite{Hikage_2019}, DES~\cite{Abbott:2017wau} and KiDS+Viking~\cite{Hildebrandt:2018yau}. In particular, the recent joint analysis of KIDS1000+BOSS+2dfLenS has yielded $S_8 = 0.766^{+0.020}_{-0.014}$ \cite{Heymans:2020gsg}, in $\sim 3\sigma$ discrepancy with $\Lambda$CDM from {\em Planck}. While an unknown systematic effect at the origin of this discrepancy is not excluded, the existence of several independent observations disfavoring the $\Lambda$CDM predictions strengthen the case for new physics.
In this article, we show that the $S_8$ tension can be resolved if DM experiences 2-body decays where the decay products are one massive warm DM (WDM) particle and one (massless) dark radiation (DR) component. We will refer to the full model as $\Lambda$DDM. We find that it requires decaying cold DM (DCDM) to have a lifetime of  $\Gamma^{-1}  \simeq  55 \ \text{Gyrs} $ and to transfer a fraction $\varepsilon\simeq 0.7 \ \%$ of its rest mass energy into the DR species. Interestingly, depending on the velocity-kick received by the massive daughter, this scenario could help resolving some of the sub-galactic scales issues in $\Lambda$CDM (e.g. Refs.~\cite{LinEtAl2001,SanchezSalcedo2003,CembranosEtAl2005,Kaplinghat2005,StrigariEtAl2007c,BorzumatiEtAl2008,PeterEtAl2010a,PeterEtAl2010,WangEtAl2014}) or could explain the recent Xenon-1T excess \cite{Xenon1tEtAl2020,KannikeEtAl2020}. 

Many authors have attempted to explain the $S_8$ tension through new properties of DM (see,~e.g.,~\cite{Lesgourgues:2015wza,Berezhiani:2015yta,Chacko:2016kgg,Murgia:2016ccp,Kobayashi:2017jcf,DiValentino:2017rcr,Buen-Abad:2017gxg,Raveri:2017jto,Poulin:2018cxd,Lin:2018nxe,DEramo:2018vss,Dutta:2018vmq,Archidiacono:2019wdp,vattis_late_2019,Bohr:2020yoe,Heimersheim:2020aoc}).  Scenarios where the DM decays only into DR have been discussed in this context, but have been shown to be at odds with the latest Planck CMB lensing and BAO data \cite{Aoyama:2014tga,Enqvist:2015ara,Chudaykin:2016xhx,Poulin:2016nat,Bringmann:2018jpr,Pandey:2019plg}.
The extension of these studies to the case of a massive daughter was recently performed in Ref.~\cite{vattis_late_2019}, where it was suggested that this scenario could resolve the `Hubble tension' -- This refers to the $\sim 5\sigma $ discrepancy between the value of the current expansion rate of the universe inferred from {\em Planck} data~\cite{Cosmo_collaboration2018planck} under $\Lambda$CDM, and that measured using the cosmic distance ladder~\cite{2019NatRP...2...10R,Verde:2019ivm, Aylor:2018drw, Wong:2019kwg,Riess:2020fzl,Soltis:2020gpl}, although not all measurements show such discrepancy \cite{Freedman:2019jwv,Freedman:2020dne,Cerny:2020inj}. However, a recent series of analysis has shown that a combination of  BAO, uncalibrated SNIa \cite{Haridasu:2020xaa} and Planck data \cite{Clark:2020miy} excludes this model. Yet, authors of Refs.~\cite{vattis_late_2019,Haridasu:2020xaa,Clark:2020miy} have limited their analyses to the $\Lambda$DDM background evolution, or neglected the perturbations of the massive daughter particles.

In this work, we perform the first thorough analysis of the $\Lambda$DDM including a realistic treatment of linear cosmological perturbations. Our careful treatment of the warm daughter perturbations allows us to pin down the space of parameters resolving the $S_8$-tension. Indeed, the warm component produced by decay leads to a suppression of the matter power spectrum at late times, similar to that of massive neutrinos or standard WDM. However, contrarily to the latter scenarios, the specific time-dependence of the power suppression imprinted by the decay allows to accommodate  Planck, BAO, uncalibrated SN1a and S8 measurements, though we confirm that it cannot simultaneously resolve the Hubble tension.  Finally, we briefly discuss implications of these results for DM model building, Xenon-1T and the `small-scale crisis of CDM'.

\section{Cosmology of 2-body DCDM} 
Our framework is characterized by two additional free parameters with respect to $\Lambda$CDM: the DCDM lifetime, $\Gamma^{-1}$, and the fraction of DCDM rest mass energy converted into DR $\eps= (1/2)[1-m^2_\wdm/m^2_\dcdm]$, where $0 \leq \eps \leq 1/2$. $\eps = 0$ corresponds to the standard CDM case (no decay),  $\eps = 1/2$ to the DM decaying solely into DR.
The general form of the background and linear perturbation equations for the DCDM, WDM and DR components can be found in Ref.~\cite{Wang:2012eka,Aoyama:2014tga}.  
We reproduce them in appendix to make this article self-contained.

The dynamics of linear perturbations in the DR
component is greatly simplified by integrating its phase-space distribution (PSD)  over all the momentum degrees of freedom~\cite{Poulin:2016nat} (after expanding it over Legendre polynomials).
To compute the WDM dynamics one cannot apply the same strategy, since the integral over the phase space is not analytic.  
One must follow the evolution of the full time- and momentum-dependent PSD, which requires solving $\mathcal{O}(10^8)$ linear differential equations for realistic cosmological analyses.
We tackle this issue by devising a new fluid approximation for the WDM species,  based on the treatment of massive neutrinos as a viscous fluid by Refs.~\cite{Lesgourgues:2011rh,Blas_2011}, which integrates out the dependency on momenta and reduces the hierarchy of equations to the first three multipoles on sub-Hubble scales. All relevant equations, and details about our fluid approximation, are given in the appendix, and derived in Ref.~\cite{Abellan:2021bpx}.
Additionally, we show therein that our approximation is accurate at the ${\cal O}(1\%)$ level in the matter power spectrum, and at ${\cal O}(0.1\%)$ precision in the CMB power spectra.

\begin{figure}[b!]
    \centering
    \includegraphics[scale=0.43]{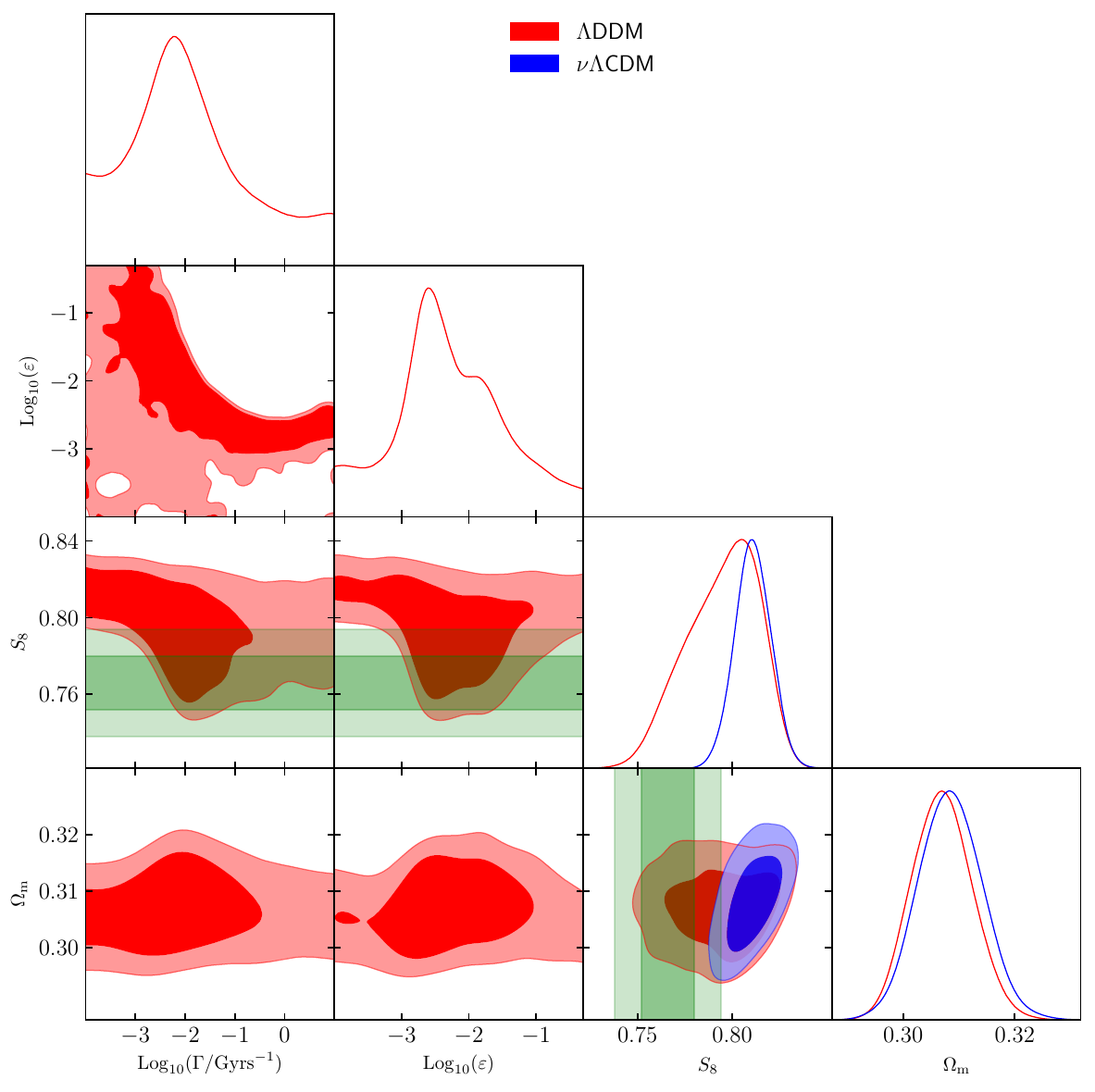}
    \caption{Reconstructed 2D posterior distributions of a subset of parameters in the $\Lambda$DDM and in the $\nu\Lambda$CDM models when confronted to Planck CMB data \cite{Cosmo_collaboration2018planck}, BOSS BAO data \cite{Beutler:2011hx,Ross:2014qpa,Alam:2016hwk}, eBOSS DR14 Ly-$\alpha$~\cite{Agathe:2019vsu, Blomqvist:2019rah}, the Pantheon SNIa catalog~\cite{Scolnic:2017caz} and a prior on $S_8=0.766^{+0.02}_{-0.014}$  \cite{Heymans:2020gsg} (green band).} 
    \label{fig:triangle_plots}
\end{figure}

\section{Resolving the $S_8$ tension with DCDM} The 2-body $\Lambda$DDM scenario under study is fully described by the following set of parameters: 
$\left\{\Omega_b h^2, \text{ln} \left(10^{10} A_s\right), n_s, \tau_{\rm reio},  \Odcdm, H_0, {\mathrm{Log}_{10}}~\Gamma ,{\mathrm{Log}_{10}}~\eps \right\}$. We implement the DDM equations in the publicly available numerical code \texttt{CLASS} \cite{Lesgourgues:2011rh,Blas_2011},  as described  in App.~\ref{sec:fluid_approx}. We use a shooting method to satisfy the budget equation. We  make use of the code \texttt{MONTEPYTHON-v3}~\cite{Audren:2012wb,Brinckmann:2018cvx} to perform a Monte Carlo Markov chain (MCMC) analysis with a Metropolis-Hasting algorithm, testing the $\Lambda$DDM model against the high-$\ell$ CMB TT, TE, EE `lite'+low-$\ell$ TT,EE+lensing data from Planck~\cite{Cosmo_collaboration2018planck}, BAO data from 6dF~\cite{Beutler:2011hx}, SDSS DR7~\cite{Ross:2014qpa}, BOSS DR12 (including $f\sigma_8$ measurements)~\cite{Alam:2016hwk}, eBOSS DR14 Ly-$\alpha$~\cite{Agathe:2019vsu, Blomqvist:2019rah} and the Pantheon SNIa catalog~\cite{Scolnic:2017caz}. Because the full KIDS1000+BOSS+2dfLenS likelihood in not yet available, we model it with a split-normal function on $S_8=0.766^{+0.02}_{-0.014}$  \cite{Heymans:2020gsg}. We expect it to be a good approximation given that the KiDS collaboration concluded that their $S_8$ value is only weakly sensitive to the effect of non-zero neutrino masses \cite{Joudaki:2019pmv}, a model with features very similar to the $\Lambda$DDM model.  (Making use of $S_8 = 0.755^{+0.019}_{-0.021}$ from KiDS450+DESY1 \cite{Joudaki:2019pmv} yields similar results.)  Note that we neglect the potential co-variance between $S_8$ and BOSS BAO/FS data for simplicity, and we checked that removing the BAO/FS data from the analysis does not affect the result.
We adopt the uninformative priors $-4 \leq {\mathrm{Log}_{10}}(\eps) \leq {\mathrm{Log}_{10}}(0.5)$, $ -4 \leq {\mathrm{Log}_{10}}(\Gamma/[{\rm Gyr^{-1}}]) \leq 1$ and $0 \leq \Odcdm \leq 1$. To gauge the importance of the late-time decay in the success of the solution, we compare the $\Lambda$DDM model with another cosmological scenario that features a power suppression at small scales, namely massive neutrinos ($\nu\Lambda$CDM). We model these as three degenerate states and vary the total mass $M_\nu$, on top of the standard $\Lambda$CDM parameters.  We assess the remaining level of tension by computing the $Q_{\rm DMAP}$ (for ``difference in the maximum a posteriori'') tension metric introduced in Ref.~\cite{Raveri:2018wln}, which essentially (for flat priors) makes use of the difference in $\chi^2$ between the fit of a given model with and without including the $S_8$ data point. 
The tension is then estimated as $\sqrt{\Delta\chi^2}$ in unit of $\sigma$.
Finally, we also compute the Bayesian evidence with the sampler {\sc MultiNest} \cite{Feroz:2008xx}, taking 1000 live points and a tolerance condition on the evidence for stopping the sampling equal to 0.1. We perform model comparison by calculating $\Delta{\rm log} B = {\rm log} B(\Lambda{\rm DDM})-{\rm log} B(\Lambda{\rm CDM})$.

Our results are reported in Tab.~\ref{table:bestfit} and summarized in Fig.~\ref{fig:triangle_plots}: In the $\Lambda$DDM scenario (red contours) we find that the best-fit (when including the $S_8$ prior) has $\eps\simeq 0.7\%$ and $\Gamma^{-1}\simeq 55$ Gyrs, yielding $S_8 \simeq 0.767$ and $\Omega_m\simeq 0.31$, in excellent agreement with the KiDS1000+BOSS+2dfLenS measurement. Moreover, the decrease in $S_8$ is driven by a smaller $\sigma_8$, while $\Omega_m$ is not affected, which is also what is favored by the data. We find a strong negative correlation between $\varepsilon$ and $\Gamma$, which approximately scales like $\Gamma^{-1} \simeq 55~ (\varepsilon/0.007)^{1.4}$ Gyrs. On the other hand, the $\nu\Lambda$CDM model can only achieve $S_8 \simeq 0.81$, with $M_\nu< 0.1614 \ \rm{eV}$ (95\% C.L.). Remarkably, we find a $\Delta\chi^2_{\rm min}\equiv\chi^2_{\rm min}(\Lambda{\rm DDM})-\chi^2_{\rm min}(\nu\Lambda{\rm CDM} )\simeq -5.5$ in favor of the $\Lambda{\rm DDM}$ model. The negative $\Delta\chi^2$ is driven  entirely by the low $S_8$ value  (The $\chi^2_{\rm min}$ per experiment is reported in App.~\ref{app:app_chi2}). The fit to other data set is barely affected by the inclusion of a $S_8$ prior in the $\Lambda$DDM model (but degrades in the $\nu\Lambda$CDM case), such that without the $S_8$ prior, there is no preference for DDM and $S_8$ seems unchanged.  Looking at the reconstructed $S_8=0.821_{-0.011}^{+0.017}$ in the $\Lambda$DDM model without the prior information, one might naively expect $S_8\simeq0.767$ to be largely excluded. In fact, the combined $\chi^2$ with the prior on $S_8$ only increases by $\sim+1.6$ (as opposed to $\sim+7.1$ in the $\nu\Lambda$CDM). This is because $S_8$ has a non-Gaussian posterior with a tail extending to low values due the degeneracy between $\Gamma$ and $\varepsilon$. This degeneracy becomes clear when incorporating the $S_8$ prior. As a result, the $Q_{\rm DMAP}$ estimator indicates  that the tension evolves from $2.7\sigma$ within $\Lambda$CDM to $1.3\sigma$ within $\Lambda$DDM.
Nevertheless, the model comparison is slightly in favor of $\Lambda$CDM,  $\Delta{\rm log} B = -0.81$,  although based on the modified Jeffrey's scales \cite{Jeffreys61,Nesseris:2012cq,Trotta:2008qt} the preference is `weak' or `inconclusive'. We thus conclude that, while the tension between our baseline data set and $S_8$ is resolved in the $\Lambda$DDM, current data do not favor the model in a Bayesian sense\footnote{ Let us point out that, assuming the $\Delta\chi^2$ is $\chi^2$-distributed with 2 degrees of freedom, the $\Lambda$DDM model is favored at 93.3\% ($\sim 2 \sigma$) over $\Lambda$CDM (in the combined analysis). This indicates that part of the `inconclusive' evidence is driven by our choice of wide priors and that different choices can affect the Bayesian evidence.}.	
Finally, we checked that the $\Lambda$CDM model with two massless neutrinos and one massive with $M_{\nu}\!=\!0.06$ eV  yields results very similar to the $\nu\Lambda$CDM model. Similarly, letting the neutrino masses free to vary in the  $\Lambda$DDM model does not affect the results (see App.~\ref{app:mnu}).  We also note that making use of linear priors on $\eps$ and $\Gamma$ does not affect the reconstructed $S_8$ value, but the scale chosen for the prior (i.e. sampling over the original prior range or in a more restricted range where $\varepsilon \sim \mathcal{O}(10^{-2})$) affects the reconstructed 2D posteriors of $\eps$ and $\Gamma$. We discuss these issues in more details in App.~\ref{app:linear_prior}.

\begin{figure}
    \centering
    \includegraphics[scale=0.28]{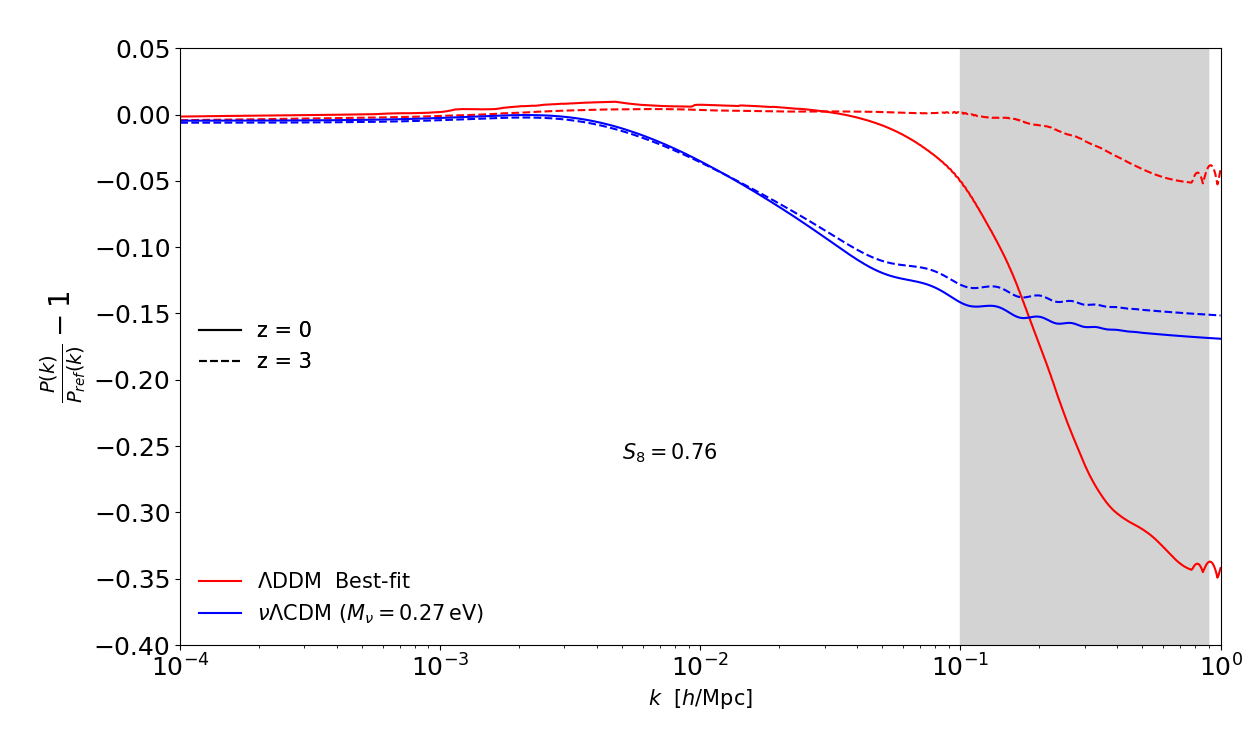}
    \includegraphics[scale=0.28]{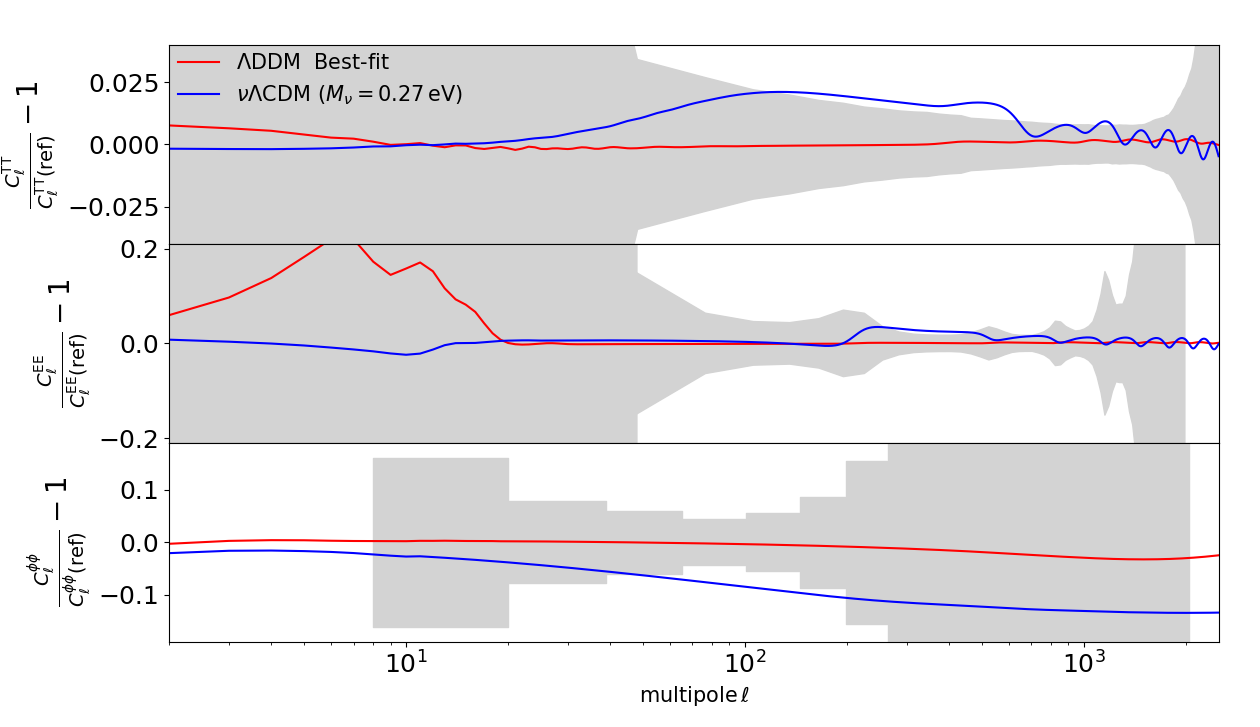}
    \caption{ \small {\em Top panel} $-$ Residuals in the linear matter power spectrum $P(k)$ at redshifts $z= 0, 3$, with respect to our baseline  $\nu\Lambda$CDM model,
    for the best-fit $\Lambda$DDM model (red lines) and a $\nu\Lambda$CDM scenario yielding the same $S_8$ (blue lines). The gray band indicates the approximate range of comoving wavenumbers contributing to $\sigma_8$.
    {\em Bottom panel} $-$ Same as above, for the (lensed) CMB TT, EE and lensing power spectrum. In this case, the gray bands show \textit{Planck} 1$\sigma$ errors. }
    \label{fig:spectra}
\end{figure}

\section{Best-fit cosmology and impact on CMB and LSS$-$} A C+WDM scenario like this one is expected to produce a suppression in the linear matter power spectrum on intermediate and small scales, with a non-trivial shape~\cite{Murgia:2017lwo,Murgia:2018now,Miller:2019pss,Bohr:2020yoe}. The cut-off scale, as in the case of massive neutrinos, is determined by the WDM free-streaming scale $k_{\rm fs}$. At wavenumbers $k > k_{\rm fs}$, pressure becomes important and structure formation is inhibited. To better understand the $\Lambda$DDM success in resolving the $S_8$ tension compared to the case of massive neutrinos, we now illustrate the impact of both models on the relevant cosmological observables for our study. 

\begin{figure}[b]
\centering
\includegraphics[scale=0.31]{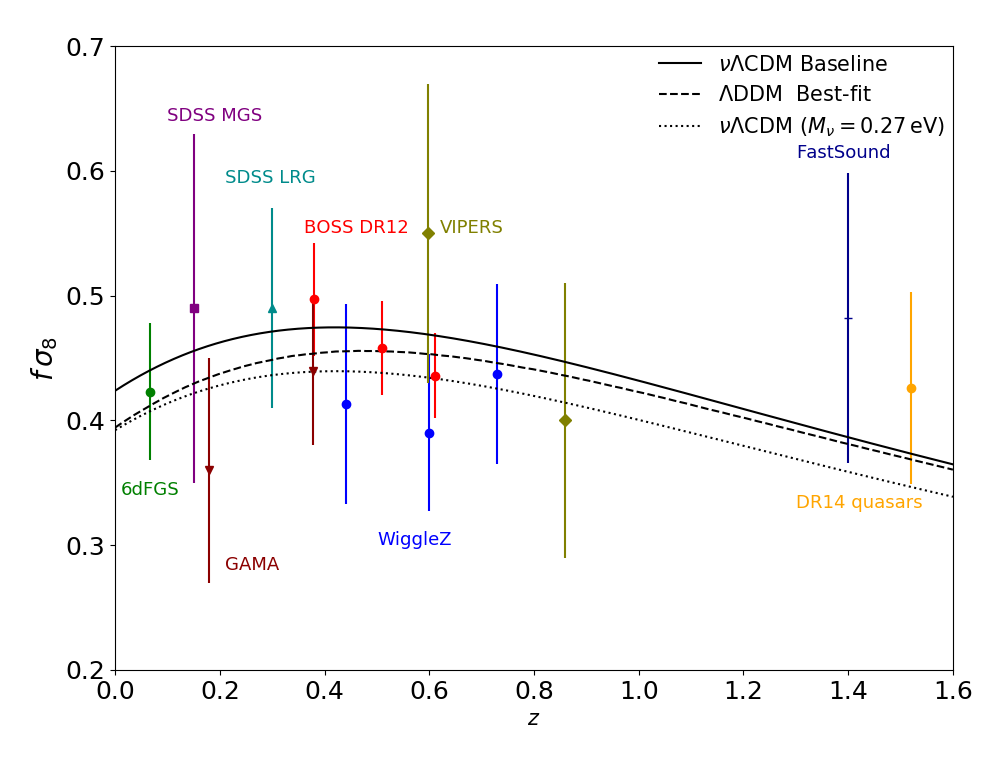}
\caption{\small Growth rate of matter fluctuations for our baseline $\nu\Lambda$CDM model (solid line),
compared to the best-fit $\Lambda$DDM model (dashed line) and to the $\nu\Lambda$CDM scenario yielding the same $\sigma_8$ and $\Omega_{\rm m}$ (dotted line). The observational constraints are taken from Ref.~\cite{Cosmo_collaboration2018planck} and references therein.}
\label{fig:fs8}
\end{figure}

In the top panel of Fig.~\ref{fig:spectra} we compare the residual differences in linear matter power spectrum $P(k)$ with respect to our baseline $\nu\Lambda$CDM model (first column of Tab.~\ref{table:bestfit}), for both the best-fit $\Lambda$DDM scenario (fourth column of Tab.~\ref{table:bestfit}) and a $\nu\Lambda$CDM model with three degenerate massive neutrinos of total mass $M_{\nu}=0.27$ eV (we adjust $\omega_{\rm cdm}= 0.1154$ whereas all other parameters are fixed to the baseline $\nu\Lambda$CDM model),  which yields $\sigma_8 \simeq 0.75$ and $\Omega_{\rm m} \simeq 0.31$, in agreement with weak lensing data \cite{Heymans:2020gsg}.
These scenarios feature two key differences: i) a distinct redshift evolution for the power suppression. In the $\Lambda$DDM scenario, it is less significant at higher redshifts, since the abundance of the WDM daughter is smaller; ii) a time-evolving cut-off scale; in the $\Lambda$DDM model, $k_{\rm fs}= \sqrt{3/2} \mathcal{H}(a)/c_g(a)$, while in the $\nu\Lambda$CDM it is obtained by evaluating $k_{\rm fs}$ at the redshift at which neutrinos become non-relativistic \cite{Lesgourgues:2015wza}. As a consequence, the CMB power spectra, well constrained by {\em Planck}, are vastly different. This is illustrated in Fig.~\ref{fig:spectra} bottom panel, for both the best-fit $\Lambda$DDM scenario and the $\nu\Lambda$CDM model which yields the same $S_8$ value. The $\nu\Lambda$CDM predicts different early-integrated Sachs-Wolfe effects, as well as different amount of lensing, because of a significant power suppression at $z\sim2-3$, where the CMB lensing kernel peaks \cite{Abazajian:2016yjj}. On the other hand, the differences between $\Lambda$CDM and $\Lambda$DDM until $z\sim2$ are very small, explaining why {\em Planck} cannot disentangle between the two scenarios. Detecting the DDM through its impact on CMB power spectra will be challenging, although CMB lensing measurements accurate at the $\sim1\%$ level could help (e.g.~with CMB-S4 \cite{Abazajian:2016yjj}).  We address the reader to Ref.~\cite{Abellan:2021bpx}, that contains an analysis where a DDM signal is artificially imprinted in a set of mock CMB data, explicitly showing that such a signal -- while being below the sensitivity of current CMB surveys -- would instead be captured by CMB-S4.
Furthermore, the differences between the growth rate $f\sigma_8$ in $\Lambda$DDM scenario and $\Lambda$CDM (shown in Fig.~\ref{fig:fs8}) at $0\lesssim z\lesssim1$, while below the sensitivity of current experiments measuring, could be measured by upcoming surveys such as  Euclid \cite{Amendola:2016saw}, LSST \cite{Mandelbaum:2018ouv}, and DESI \cite{Aghamousa:2016zmz}. 

\section{Some implications of the DCDM} 
Concrete realizations of $\Lambda$DDM scenarios as the ones considered in this work may arise for instance
in the context of the `super weakly interacting massive particle' (superWIMP) class of exotic particle
physics models~\cite{CoviEtAl1999,FengEtAl2003,FengEtAl2003b}, whose super weak
couplings make them evade many observational constraints \cite{AllahverdiEtAl2015,Feng2010,Choi:2021uhy}.
The decaying particle must have properties similar to CDM candidates, which sets a lower mass bound of $m\gtrsim 5$~keV if it is thermally produced in the early universe
\cite{VielEtAl2005,BoyarskyEtAl2009,IrsicEtAl2017,YecheEtAl2017,Schneider2018}, raising to the
MeV mass scale depending on couplings to standard model particles \cite{BoehmEtAl2013a}
(irrelevant for thermal production in hidden sectors \cite{FengEtAl2008d}). For non-thermally produced fermions, the Tremaine-Gunn limit \cite{TremaineEtAl1979} of $m\gtrsim 1$~keV
\cite{BoyarskyEtAl2009b} applies.

Interestingly, late decays of CDM to WDM are among the proposed cures to some
observational discrepancies with CDM on small (sub-galactic) scales after structure formation
(e.g.~\cite{LinEtAl2001,SanchezSalcedo2003,CembranosEtAl2005,Kaplinghat2005,StrigariEtAl2007c,BorzumatiEtAl2008,PeterEtAl2010a,PeterEtAl2010}, and e.g.~\cite{deBlok2010,BullockEtAl2017,PontzenEtAl2014,ReadEtAl2019,OstrikerEtAl2019,ZavalaEtAl2019a} for reviews 
on small-scale issues). A detailed inspection of the effective parameter space was performed in
Ref.~\cite{Peter2010a}, which was later supplemented by dedicated cosmological simulations
\cite{WangEtAl2014}, showing that this scenario mostly affects galaxy satellite/subhalo
properties. In particular, a daughter particle speed $3\times 10^{-5} \lesssim v\simeq\eps \lesssim
2\times 10^{-3}$ may reduce the abundance of subhalos and their concentrations for a large range of
lifetimes, up to $\sim 100$~Gyr. Greater speeds are disfavored from the existence of dwarf
galaxies (see also~\cite{WangEtAl2013}), unless $\Gamma^{-1}\gtrsim 100$~Gyrs, for which 
$\Lambda$DDM does not depart from $\Lambda$CDM as far as structure formation is concerned (this 
also holds for very low speeds irrespective of $\Gamma$).
Further constraining our favored range for  $\Lambda$DDM models along those lines would require to
predict the non-linear matter power spectrum, which goes beyond the scope of this paper (An improved mildly non-linear computation could also benefit the modeling of the lensing signal, including CMB lensing.). However, our best-fitting parameters do fall in ranges that could partly address small-scale CDM challenges.

Remarking from our results that the 1-$\sigma$ range for the speed of the daughter particle extends to $v \simeq \varepsilon \sim 0.05$, brings up the intriguing possibility that this scenario may be connected with the excess events in the electronic recoils recently reported by the Xenon-1t Collaboration \cite{Xenon1tEtAl2020}. This excess, if interpreted in terms of elastic interactions of DM with electrons, would point to DM particles of mass $m \gtrsim 1$~MeV with unexpected high speeds of $v\sim 0.1$ \cite{KannikeEtAl2020,Xu:2020qsy}. In our scenario, this would set the mass scale for the parent CDM particle $m_{\rm dcdm}=m_{\rm wdm}/\sqrt{1-2\varepsilon}$. This can be achieved in both thermal and non-thermal production scenarios for the DCDM particle, with the latter favored for $m_{\rm dcdm}\lesssim 10$~MeV to prevent it from being warm. The contact interaction cross section of the warm daughter with electrons is constrained to be $\sigma_e \sim 10^{-45}{\rm cm^{2}}(f_{\rm wdm}/0.1)^{-1}(\rho_\odot/0.4\,{\rm GeV/cm^3})^{-1}(m_{\rm wdm}/1\,{\rm MeV})$ \cite{KannikeEtAl2020} (even lower for pseudo-scalar contact interactions \cite{Buttazzo:2020vfs}), and should be mediated by an independent scalar sector to ensure that the WDM particle is not produced in the early universe.
Alternatively, the massive daughter particle might be a dark photon with kinetic mixing with photons, which could then deposit its entire mass energy onto electrons from a photoelectric-like process \cite{PospelovEtAl2008a}. This sets $m_{\wdm}\sim 2$~keV and no longer relies on $\varepsilon$, which can therefore only be achieved in non-thermal production scenarios. A possible effective Lagrangian for this scenario could be ${\cal L}_{\rm eff}\supset \tilde{e}\gamma_\mu\bar\chi \tilde{A}^\mu\xi-(\kappa/2)F_{\mu\nu}{\cal F}^{\mu\nu}$, with $\chi$, $\xi$, $\tilde{A}_\mu$, and ${\cal F}^{\mu\nu}$ being the DM particle (a fermion here), its daughter (possibly a massless Weyl fermion), a dark $U(1)$ gauge field (possibly massive), and its associated strength field, which may find realizations e.g. in models with a broken $U(1)_{\text B-L}$ gauge symmetry, also proposed, incidentally, to solve small-scale CDM issues (e.g.~\cite{ChoiEtAl2020,ChoiEtAl2020a,ChoiEtAl2020b}).

\begin{table*}[t!]
\centering
\bgroup
\def\arraystretch{1.2}
\setlength\tabcolsep{1mm}
 \begin{tabular}{|l|c|c|c|c|} 
 \hline
 Model & \multicolumn{2}{c|}{$\nu\Lambda$CDM} & \multicolumn{2}{c|}{$\Lambda$DDM} \\ [0.5ex]
 \hline
 \hline
  Parameter & w/o $S_8$ & w/ $S_8$ & w/o $S_8$ & w/ $S_8$ \\ [0.5ex] 
 \hline
$100 \ \omega_b $  & $2.245(2.242) \pm 0.013$  & $2.251 (2.253) \pm 0.013 $ & $2.243(2.244)_{-0.013}^{+0.014}$  & $2.246(2.241) \pm 0.013$    \\ 
$  \omega_{\rm cdm} \ \text{or} \ \omega_{\rm dcdm}^{\rm ini}$ & $0.1193(0.1194) \pm 0.0009$ & $0.1182(0.1184)_{-0.0008}^{+0.0009}$ & $0.1195(0.1195)\pm0.00095$ &  $0.1191(0.1194)_{-0.001}^{+0.0009}$ \\ 
$ H_0 /[{\rm km/s/Mpc}]$ & $67.55(67.76)_{-0.44}^{+0.46}$& $67.85(68.08)_{-0.44}^{+0.47}$ & $67.71(67.71)\pm0.42 $ &  $67.92(67.70)_{-0.42}^{+0.43}$   \\
$\text{Ln}(10^{10} A_s)$ & $3.052(3.045)_{-0.016}^{+0.014}$ & $3.047(3.043)_{-0.015}^{+0.014}$  &  $3.051(3.052)_{-0.015}^{+0.014}$&  $3.048(3.052)_{-0.016}^{+0.014}$  \\
$ n_s$ & $0.9676(0.9663) \pm 0.0037$ & $0.9697(0.9683)_{-0.0036}^{+0.0037}$ &$0.9674(0.9672) \pm 0.0038 $  &  $0.9682(0.9673) \pm 0.0037 $\\
$\tau_{\rm reio}$ & $0.058(0.055)_{-0.008}^{+0.007}$ & $0.0569(0.0549)_{-0.008}^{+0.007}$ & $0.0576(0.0582)_{-0.0076}^{+0.0071}$ &  $0.0570(0.0582)_{-0.0077}^{+0.0071}$   \\
$ M_{\nu} / \text{eV} $ & $<0.1395  $ & $ < 0.1611 $ &  $-$ & $-$ \\ 
$\text{Log}_{10} (\varepsilon)$ & $-$ &$-$ &  $-2.69(-2.97)_{-1.3}^{+0.32}$ & $-2.28(-2.16)_{-0.78}^{+0.8}$ \\ 
$\text{Log}_{10} (\Gamma/[{\rm Gyr}^{-1}])$ &$-$ & $-$ & unconstrained (-3.86) & $-1.89(-1.74)_{-1.5}^{+0.82}$ \\ 
 \hline
 $\Omega_{\rm m}$ & $0.3127(0.3104)_{-0.0061}^{+0.0057}$ & $0.3083(0.3061)_{-0.006}^{+0.0056}$ & $0.3102(0.3109)_{-0.0058}^{+0.0056}$ & $0.3071(0.3099)_{-0.0058}^{+0.0053}$  \\ 
 $S_8 $   & $0.824(0.824) \pm 0.011$& $0.81(0.816) \pm 0.01$ & $0.821(0.828)_{-0.011}^{+0.017}$ &  $0.795(0.767)_{-0.016}^{+0.024}$ \\ 
 \hline
 $\chi^2_{\rm min}$ & 2053.4 & 2060.5 & 2053.4 & 2055.0\\
 $\Delta{\rm log} B$ & 0 & 0 & -1.4 & -0.81\\
\hline
 $Q_{\rm DMAP}$ tension & \multicolumn{2}{c|}{2.7$\sigma$} & \multicolumn{2}{c|}{1.3$\sigma$}\\
 \hline
\end{tabular}
\egroup
\caption{The mean (best-fit) $\pm 1\sigma$ errors of the cosmological parameters from the analysis of Planck, BAO/FS, SN1a data, with and without a split-normal likelihood on $S_8$ from Ref.~\cite{Heymans:2020gsg}. For each model and data-set, we also report the best-fit $\chi^2$, the level of tension estimated through the  $Q_{\rm DMAP}$ metric \cite{Raveri:2018wln} and the Bayesian evidence. }
\label{table:bestfit}
\end{table*}

\section{Conclusions} In this article we have studied the cosmological phenomenology of a scenario in which CDM decays into one WDM and one DR species. Through a comprehensive MCMC analysis including up-to-date data, we demonstrated that the $\Lambda$DDM scenario can resolve the $S_8$ tension. The inferred $S_8$ value in the analysis combining all data sets is in excellent agreement with the direct measurement from WL, due to the suppression in the gravitational clustering induced by WDM free-streaming in a similar fashion to massive neutrino or standard WDM cosmologies. However contrarily to the latter, the specific time dependence of the power suppression imprinted by the decay allows to accommodate simultaneously all the data considered in this work. 
Finally, we briefly outlined implications and possible future developments of our study, in view of DM model building, the $\Lambda$CDM small-scale crisis and the recent Xenon-1T excess. 
Future experiments measuring the growth rate of fluctuations at $0\lesssim z\lesssim1$ or the CMB lensing power spectrum to $\sim1\%$ accuracy will further test this scenario. 
\\
\\

{\em Acknowledgements} The authors warmly thank Rodrigo Calderón, Zackaria Chacko, Abhish Dev, Peizhi Du, Yuhsin Tsai, Julien Lesgourgues and Pasquale Serpico for many comments and discussions.  This work has been partly supported by the CNRS-IN2P3 grant Dark21, the ANR project ANR-18-CE31-0006, the national CNRS-INSU programs PNHE and PNCG, and the European Union's Horizon 2020 research and innovation program under the Marie Sk\l{}odowska-Curie grant agreements N$^\circ$
  690575 and N$^\circ$ 674896. The authors acknowledge the use of computational resources from the Ulysses SISSA/ICTP super-computer in Trieste, the CNRS/IN2P3 Computing Centre (CC-IN2P3) in Lyon, the IN2P3/CNRS and the Dark Energy computing Center funded by the OCEVU Labex (ANR-11-LABX-0060) and the Excellence Initiative of Aix-Marseille University (A*MIDEX) of the “Investissements d’Avenir” programme.
This project has received support from the European Union’s Horizon 2020 research and innovation program under the Marie Skodowska-Curie grant agreement No 860881-HIDDeN.\\

\newpage
\bibliography{dcdm}

\appendix

\section{Appendix A: Details about the fluid approximation for the warm daughter species }
\label{sec:fluid_approx}

\begin{figure}
\centering
\includegraphics[scale=0.26]{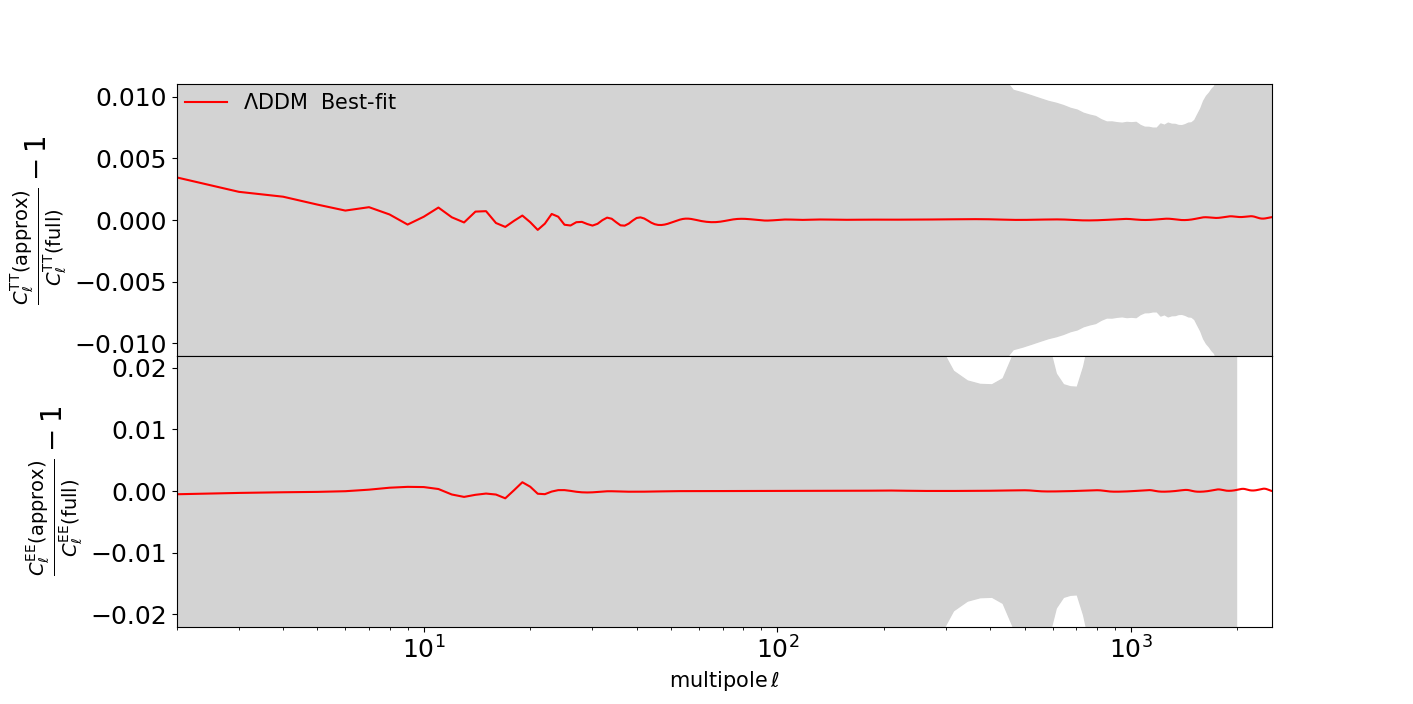}
\caption{ Residuals of the lensed CMB TT (upper) and EE (lower) power spectra computed in the fluid approximation, with respect to solving the full hierarchy of equations. Cosmological parameters are fixed to the best-fit values of the $\Lambda$DDM model. The gray bands indicate Planck 1$\sigma$ sensitivity. }
\label{fig:cl_check}
\end{figure}

\begin{figure}
\centering
\includegraphics[scale=0.31]{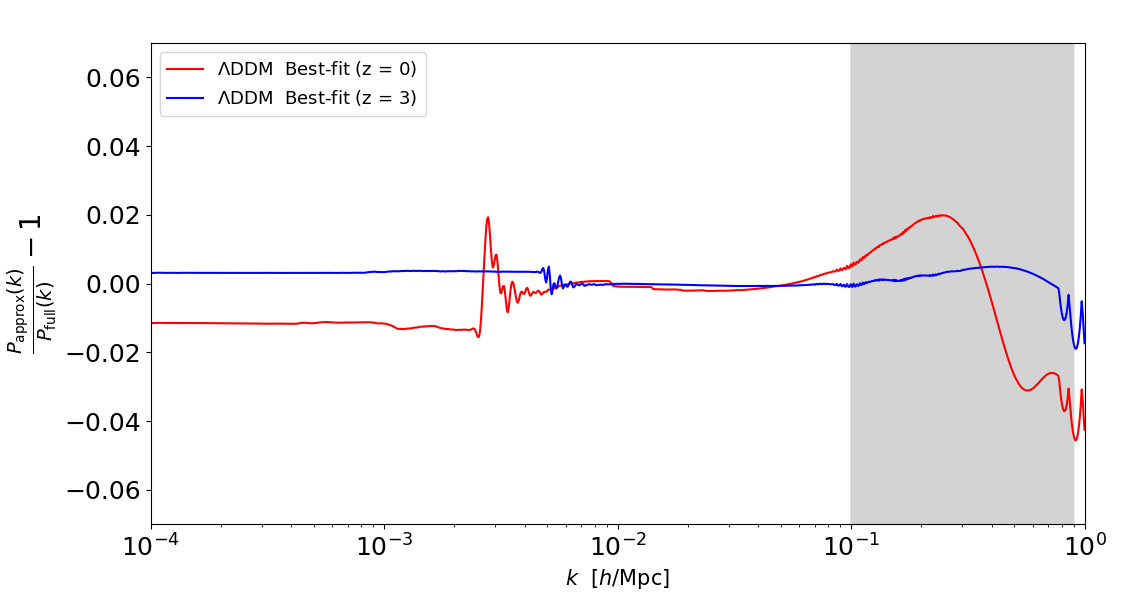}
\caption{ Residuals in the linear matter power spectrum (at redshifts $z=0, 3$) computed in the fluid approximation, with respect to solving the full hierarchy of equations. Cosmological parameters are fixed to the best-fit values of the $\Lambda$DDM model. The gray band denotes the approximate range of comoving wavenumbers contributing to $\sigma_8$. }
\label{fig:pk_check}
\end{figure}

Here we sketch the main equations that dictate the evolution of the background and perturbed phase-space distribution (p.s.d.) of the warm dark matter (for further details, see \cite{Aoyama:2014tga}). The evolution of the background p.s.d. $\bar{f}$ is given by
\begin{equation}
\frac{\partial \bar{f}}{\partial \tau} = \frac{a \Gamma \bar{N} (\tau)}{4 \pi q^2} \delta (q-a p_{\rm max}), \label{f_wdm}
\end{equation}
where $\bar{N}(\tau)$ is the mean comoving number density of the mother particle,  $\bar{N}(\tau) = (\Odcdm \rho_c /  m_\dcdm   ) \text{exp}\left(-\Gamma t\right)$,
and $p_{\rm max}$ is the initial physical momentum of the daughter particles, $p_{\rm{max}} = m_\wdm \eps/\sqrt{1-2\eps}$. The evolution of the perturbed p.s.d. multipoles $\Delta f_\ell$ is given by the following hierarchy of equations
\begin{align}
\frac{\partial}{\partial \tau} \left(\Delta f_0 \right) &= -\frac{q k}{\mathcal{E}} \Delta f_{1} 
 + q\frac{\partial \bar{f}}{\partial q}  \frac{\dot{h}}{6}
+\dot{\bar{f}} \delta_\dcdm, \label{delta_f_0}\\ 
\frac{\partial}{\partial \tau} \left(\Delta f_1 \right) &= \frac{q k}{3\mathcal{E}} \left[\Delta f_{0}-2 \Delta f_{2} \right], \label{delta_f_1}\\ 
\frac{\partial}{\partial \tau} \left(\Delta f_2 \right) &= \frac{q k}{5\mathcal{E}} \left[ 2 \Delta f_{1}-3 \Delta f_{3} \right]
 - q\frac{\partial \bar{f}}{\partial q} \frac{(\dot{h}+6\dot{\eta})}{15},   \  \label{delta_f_2} \\ 
\frac{\partial}{\partial \tau} \left(\Delta f_\ell \right) &= \frac{q k}{(2\ell+1)\mathcal{E}} \left[ \ell \Delta f_{\ell-1}-(\ell+1) \Delta f_{\ell+1} \right] \nonumber \\
&\hspace{50mm}   ( \ell \geq 3). \label{delta_f_3}
\end{align}
Here $\mathcal{E} = \sqrt{m_\wdm^2 a^2+q^2}$  is the comoving energy of the WDM, and $h$ and $\eta$ represent the usual scalar metric perturbations in the synchronous gauge. For integration of the hierarchy with respect to time, it is useful to consider $\tau_q$, which is defined as the conformal time when daughter particles with comoving momentum $q$ are born, i.e., $q = a (\tau_q) p_{\rm{max}}$. On times $\tau < \tau_q$, we set all $\Delta f_{\ell} = 0$, since no daughter particle with comoving momentum $q$ could have been produced. On the crossing time $\tau=\tau_q$, only the values of $\Delta f_{\wdm,0} (\tau_q)$ and $\Delta f_{\wdm,2} (\tau_q)$ are non-vanishing, and we set them according to the analytical formulas (A.5) and (A.7) from \cite{Aoyama:2014tga}. Finally, on times $\tau > \tau_q$, we treat the WDM component as a massive neutrino species. We choose the maximum multipole $\ell_{\rm max}$ to truncate the hierarchies of equations according to the scheme proposed in \cite{Ma:1995ey} for massive neutrinos.
\noindent
The perturbed density  of the DCDM obeys the same equation as in the CDM case \cite{Poulin:2016nat},
$\dot{\delta}_\dcdm = -\dot{h}/2$. \\ 
The mean density $\bar{\rho}_\wdm$ can be directly computed by integration of the background distribution $\bar{f}$ over the phase space:
\begin{align}
\bar{\rho}_\wdm (a) &=  \frac{  C }{a^4}\int_{0}^{a} da_q \frac{e^{-\Gamma t_q } }{\mathcal{H}_q}  \sqrt{ \eps^2 a_q^2+(1-2\eps )a^2 }, \label{rhowdm}
\end{align}
where $C \equiv \rho_c  \Odcdm \Gamma  $, $\mathcal{H}_q \equiv \mathcal{H} (a_q)$, $t_q \equiv t(a_q)$. The elements of the perturbed stress-energy tensor are computed as integrals of the first p.s.d. multipoles $\Delta f_\ell$ over momenta (we follow the convention of \cite{Lesgourgues:2011rh}). At sub-Hubble scales, and far away from the relativistic regime, we find that the dynamics of the warm daughter species is well captured if we simply consider the perturbed continuity equation
\begin{align}
\dot{\delta}_\wdm &= -3 \mathcal{H} (c^2_{\rm syn}-w_\wdm) \delta_\wdm -(1+w_\wdm) \left(\theta_\wdm+\frac{\dot{h}}{2}\right) \nonumber \\
&+a\Gamma (1-\varepsilon) \frac{\bar{\rho}_\dcdm}{\bar{\rho}_\wdm} (\delta_\dcdm-\delta_\wdm) , \label{delta_dot_wdm}
\end{align}
and the perturbed Euler equation
\begin{align}
\dot{\theta}_\wdm  &= -\mathcal{H} (1-3 c_g^2) \theta_\wdm + \frac{c^2_{\rm syn}}{1+w_\wdm}k^2\delta_\wdm -k^2 \sigma_\wdm \nonumber \\ &-a \Gamma (1-\varepsilon) \frac{1+c_g^2}{(1+w_\wdm)} \frac{\bar{\rho}_\dcdm}{\bar{\rho}_\wdm} \theta_\wdm . \label{theta_wdm}
\end{align}
We have introduced the sound speed in the synchronous gauge, $c^2_{\rm syn} \equiv \delta P_\wdm / \delta \rho_\wdm$, and the adiabatic sound speed, $c_g^2 \equiv \dot{\bar{P}}_\wdm / \dot{\bar{\rho}}_\wdm$. We find that including higher moments (such as the shear $\sigma_{\rm wdm}$) is not necessary to get an accurate description, except for the purely relativistic case $\varepsilon=0.5$. 
In practice, we set $\sigma_\wdm$ to a constant value, obtained via integration of the second PSD multipole in the Boltzmann hierarchy, just before switching on the fluid equations. 
However, contrarily to what is done in \cite{Lesgourgues:2011rh} for the fluid description of massive neutrinos, we do not assume that $c_{\rm syn}^2 \simeq c_g^2  $ for the WDM. In fact, calculations using the full Boltzmann hierarchy of Eqs. \eqref{delta_f_0}-\eqref{delta_f_3} reveal that $c_{\rm syn}^2 $ exhibits a particular $k$-dependence (namely, some small enhancement on scales $k > k_{\rm fs}$) that cannot be captured by a background quantity such as $c_g^2$. 
To address this problem, we use the following prescription 
\begin{eqnarray}
c_{\rm syn}^2 = c_g^2 \left[ 1+(1-2\varepsilon) T(k/k_{\rm fs}) \right],
\label{sync_sound_speed}
\end{eqnarray}
where $c_g^2$ is the adiabatic sound speed,
\begin{eqnarray}
c_g^2 &= w_\wdm \left(5- \frac{\mathfrak{p}_\wdm}{\bar{P}_\wdm} -\frac{\bar{\rho}_\dcdm}{\bar{\rho}_\wdm} \frac{a \Gamma}{3 w_\wdm \mathcal{H}} \frac{\varepsilon^2}{(1-\varepsilon)} \right) \nonumber \\
 &\times\left[3(1+w_\wdm)-\frac{\bar{\rho}_\dcdm}{\bar{\rho}_\wdm} \frac{a\Gamma}{\mathcal{H}} (1-\varepsilon) \right]^{-1} \label{sound_speed},
\end{eqnarray}
and $T(x) = 0.2\sqrt{x}$ is a function that was fit using the full Boltzmann hierarchy. \\
In order to check the accuracy of the fluid description for the WDM species, we compare it to the "exact" calculation of the Boltzmann hierarchy. As an example, for the cosmological parameters we take the values obtained from the best-fit analysis that includes the $S_8$ measurements. We first solve the full Boltzmann hierarchy of Eqs. \eqref{delta_f_0}-\eqref{delta_f_3}, using $10^4$ momentum bins and $l_{\rm max} = 17$. We then compare with the approximate calculation, that solves the hierarchy with only $300$ momentum bins and switches-on the fluid equations \eqref{delta_dot_wdm}-\eqref{sync_sound_speed} at sub-Hubble scales $k\tau > 25$.  We plot the residuals  between the approximate and exact calculation of the CMB and matter power spectra in Fig.~\ref{fig:cl_check} and~\ref{fig:pk_check} respectively. One can see see that the accuracy in the CMB TT and EE calculation is well below 1\%, meaning that it is safe to use the fluid approximation for CMB analysis. By looking at Fig.~\ref{fig:pk_check}, one can see that the matter power spectrum is computed at a few percent accuracy, reaching $2-3\%$ errors for the smallest scales. Nevertheless, our calculation reveals $S_8^{\rm approx} \simeq 0.767 $ and  $S_8^{\rm full} \simeq 0.762 $, i.e., a relative error of $\sim 0.65 \ \%$, which is well below the $\sim 1.8 \ \% $ relative error of the $S_8$ measurement from \cite{Heymans:2020gsg}. We thus conclude that it is safe to make use of our approximate scheme for cosmological analysis.

\section{Appendix B:  $\chi^2$ per experiment obtained in the different analysis performed}
\label{app:app_chi2}
\begin{table*}[t!]
\centering
\bgroup
\def\arraystretch{1.2}
\setlength\tabcolsep{1mm}
 \begin{tabular}{|l|c|c|c|c|} 
 \hline
 Model & \multicolumn{2}{c|}{$\nu\Lambda$CDM} & \multicolumn{2}{c|}{$\Lambda$DDM} \\ [0.5ex]
 \hline
  Experiment & w/o $S_8$ & w/ $S_8$ & w/o $S_8$ & w/ $S_8$ \\ [0.5ex] 
 \hline\hline
 \Planck~high$-\ell$ TT,TE,EE & 586.67 & 587.57 & 584.82   & 585.74  \\ 
 \Planck~ low$-\ell$ EE       & 396.06 & 396.05 & 396.92   & 396.92  \\ 
 \Planck~ low$-\ell$ TT       & 23.18 & 22.66 &  23.12   & 23.09 \\ 
 \Planck~lensing              & 8.93 &9.60  & 8.78     & 9.07 \\ 
\Pantheon                     &  1026.93& 1026.73 & 1026.94  & 1026.93  \\ 
\BAO~BOSS low$-z$             &  1.23& 1.62 & 1.20     & 1.21  \\
\BAO~\FS~BOSS DR12            &  6.51&  5.88& 6.63     & 6.95 \\ 
eBOSS DR14~Ly$-\alpha$        & 4.93 & 4.68 & 4.94     & 4.91 \\ 
\sc KIDS1000+BOSS+2dfLenS & $-$ & 5.64 & $-$      & 0.15 \\ 
 \hline
total &  2053.4& 2060.5  & 2053.4  & 2055.0  \\ 
 \hline
 \end{tabular}
\egroup
\caption{Best-fit $\chi^2$ per experiment (and total) in $\nu \Lambda$CDM and $\Lambda$DDM, with and without a split-normal likelihood on $S_8$ from Ref.~\cite{Heymans:2020gsg}. }
\label{table:chi2}
\end{table*}
\newpage
\bibliography{dcdm}

\section{Appendix C: Results with the neutrino mass free in the $\Lambda$DDM cosmology}
\label{app:mnu}
\begin{figure*}
    \centering
    \includegraphics[scale=0.4]{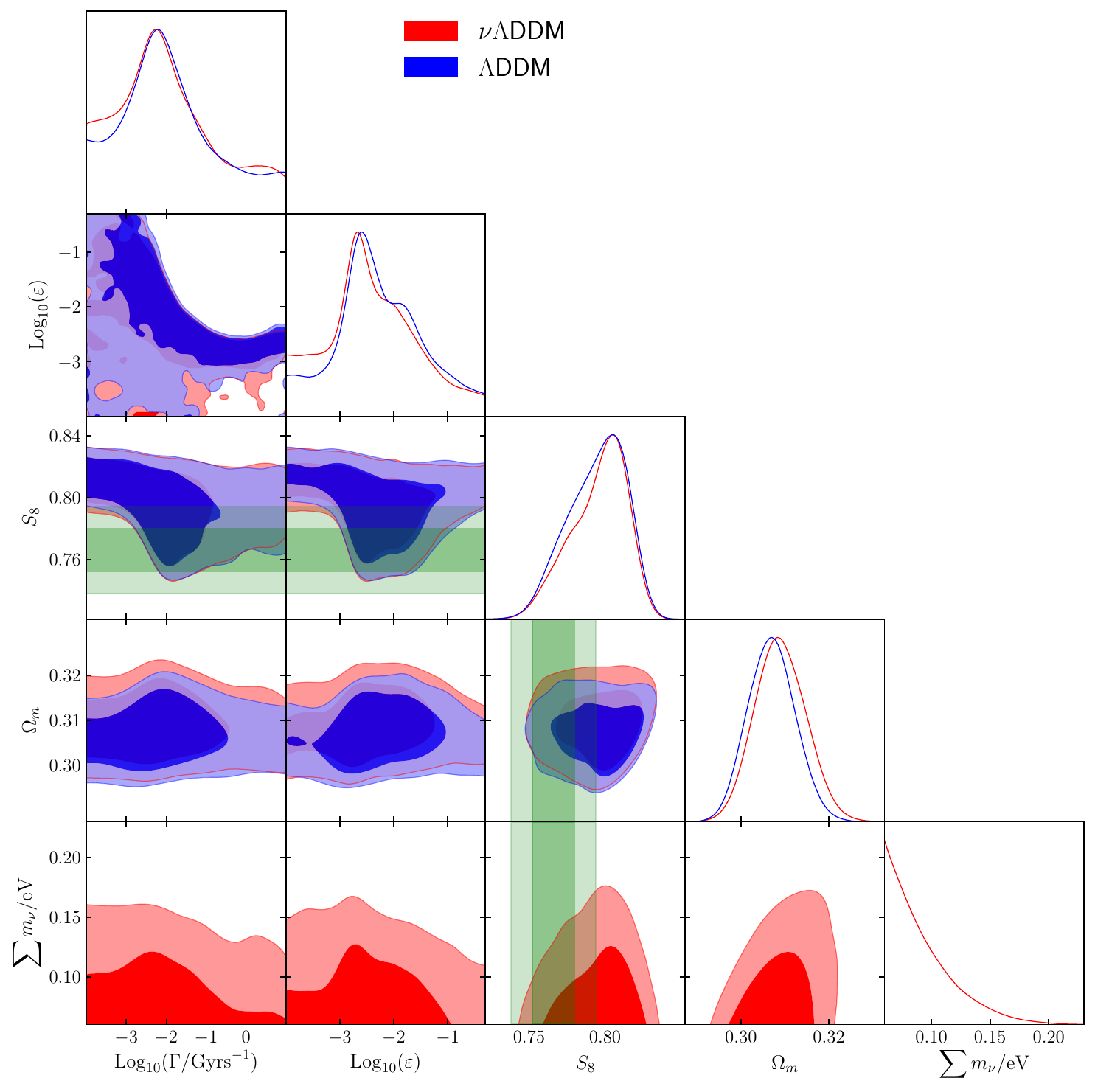}
    \caption{Reconstructed 2D posterior distributions of $\{S_8,\Omega_m,{\rm Log}_{10}(\Gamma/{\rm Gyrs}^{-1}),{\rm Log}_{10}(\varepsilon),\sum m_\nu/{\rm eV}\}$ with the neutrino mass fixed to 0.06 eV (blue) or let free to vary (red).}
    \label{fig:DDM-nu}
\end{figure*}

We show in Fig.~\ref{fig:DDM-nu} the 2D posterior distributions of $\{S_8,\Omega_m,{\rm Log}_{10}(\Gamma/{\rm Gyrs}^{-1}),{\rm Log}_{10}(\varepsilon),\sum m_\nu/{\rm eV}\}$ with the (individual) neutrino mass fixed to 0.06 eV (red) or let free to vary (blue). When considering  the neutrino mass a free parameter, we model neutrinos as three degenerate state, while when the neutrino mass is fixed to 0.06 eV we consider one massive, two massless neutrinos. One can see that the results are unaffected by our choice of keeping the neutrino mass fixed to 0.06 eV in our fiducial analysis.
\section{Results with a linear prior on $\Gamma$ and $\varepsilon$}
\label{app:linear_prior}
In our baseline analysis, we have made use of log-prior on $\varepsilon$ and $\Gamma$, to ease comparison with earlier work \cite{vattis_late_2019} who made the same choice. In this appendix, we present results using linear priors on the DDM parameters.

Now, let us stress that the use of log-prior is more agnostic, in the sense that a log-prior is a less informative prior than the linear prior. Indeed, a linear prior carries a scale (given by the large error bars used in the proposal distribution of the MCMC sampler), which makes it difficult for the sampler to detect fine structure over 4 orders of magnitude with a linear scale, especially at very small values.

Given the difference between the scale of the upper bound on $\Gamma/\rm{Gyrs}^{-1}$ an $\varepsilon$ (typically $\sim10^{-1}$) and that of the lower bound ($\sim10^{-3}$)), it is very hard to reconstruct correctly the parameter space with a linear prior.

\begin{figure*}
    \centering
    \includegraphics[scale=0.4]{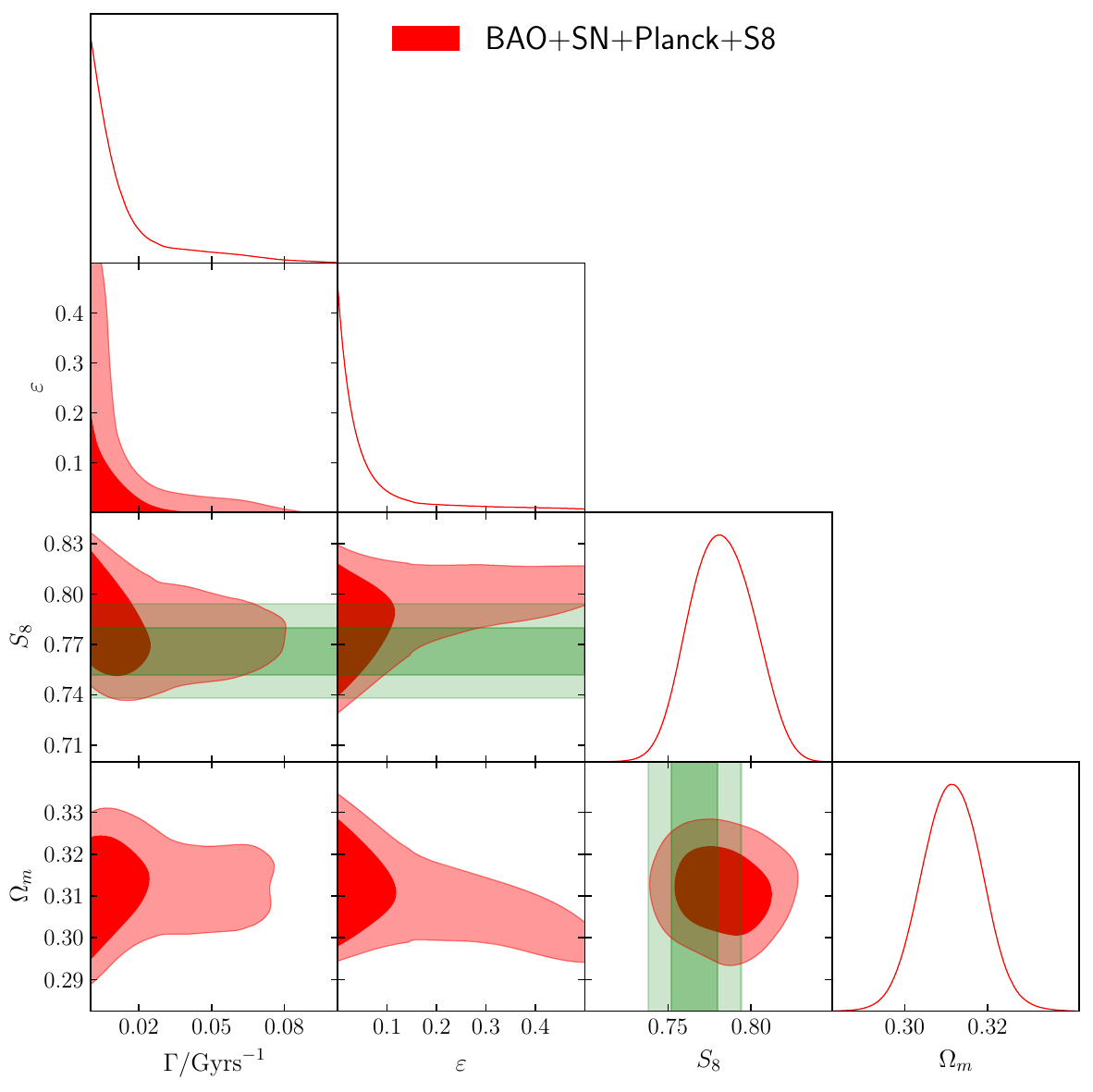}
    \includegraphics[scale=0.4]{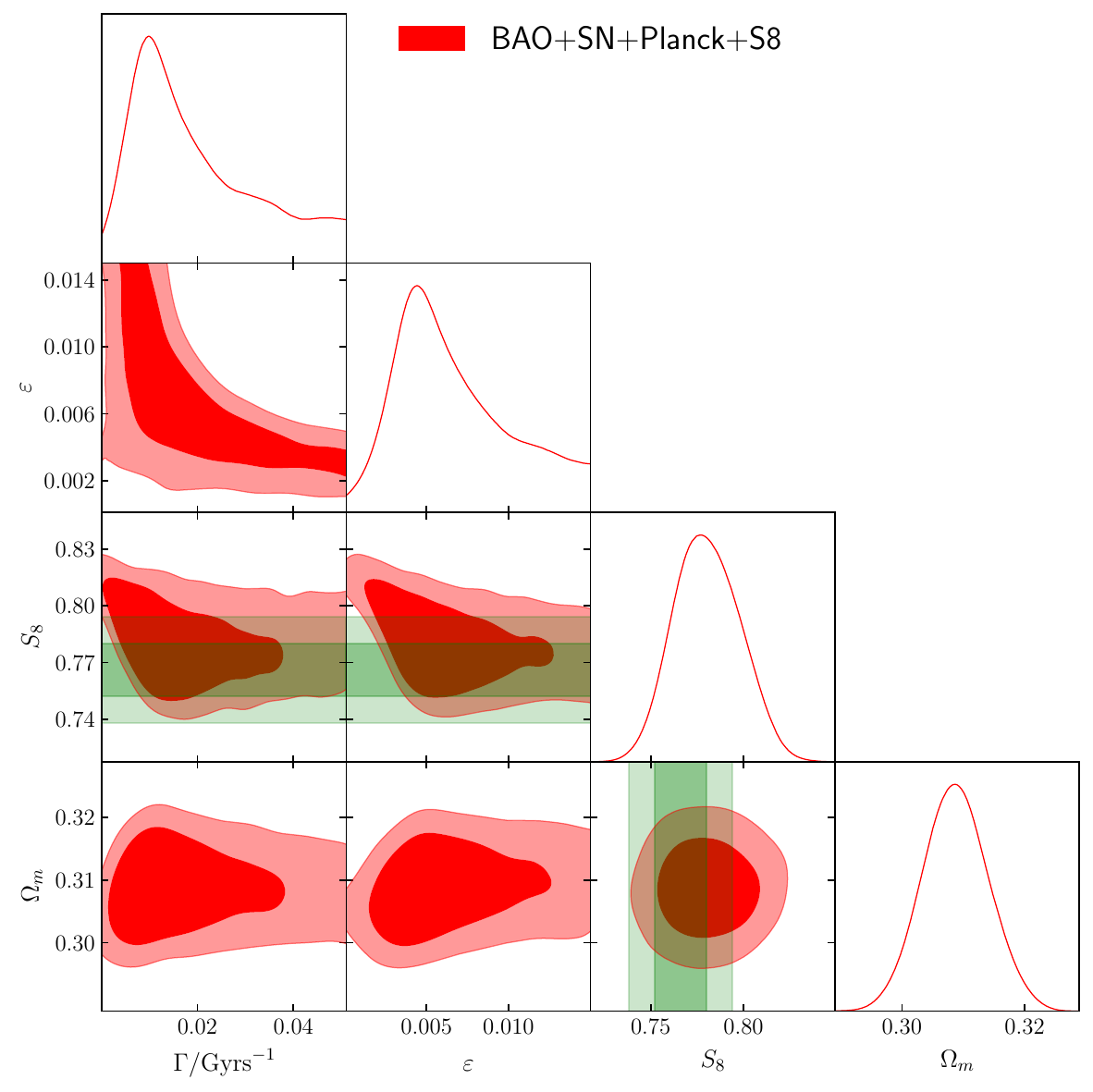}
    \caption{Reconstructed 2D posteriors with linear priors and sampling either with the original prior range (left panel) or within a restricted prior range (right panel).}
    \label{fig:linpriors}
\end{figure*}

To illustrate this difficulty we show in Fig.~\ref{fig:linpriors} the results of the linear prior analysis with two different configurations: using the original prior range, and using a more restricted range: $\varepsilon \in [0.0001,0.015]$ and $\Gamma/\rm{Gyrs}^{-1} \in [0.0001,0.05]$. In the latter case, the level of detection is un-affected: $\varepsilon$ and $\Gamma$ are non-zero at the 2$\sigma$ level, similarly to the results with the uninformative log-prior. However, in the former case, which weights in favor of larger $\varepsilon$ values, one would (wrongly) deduce an upper limit only.

\begin{figure*}
    \centering
    \includegraphics[width=1\columnwidth]{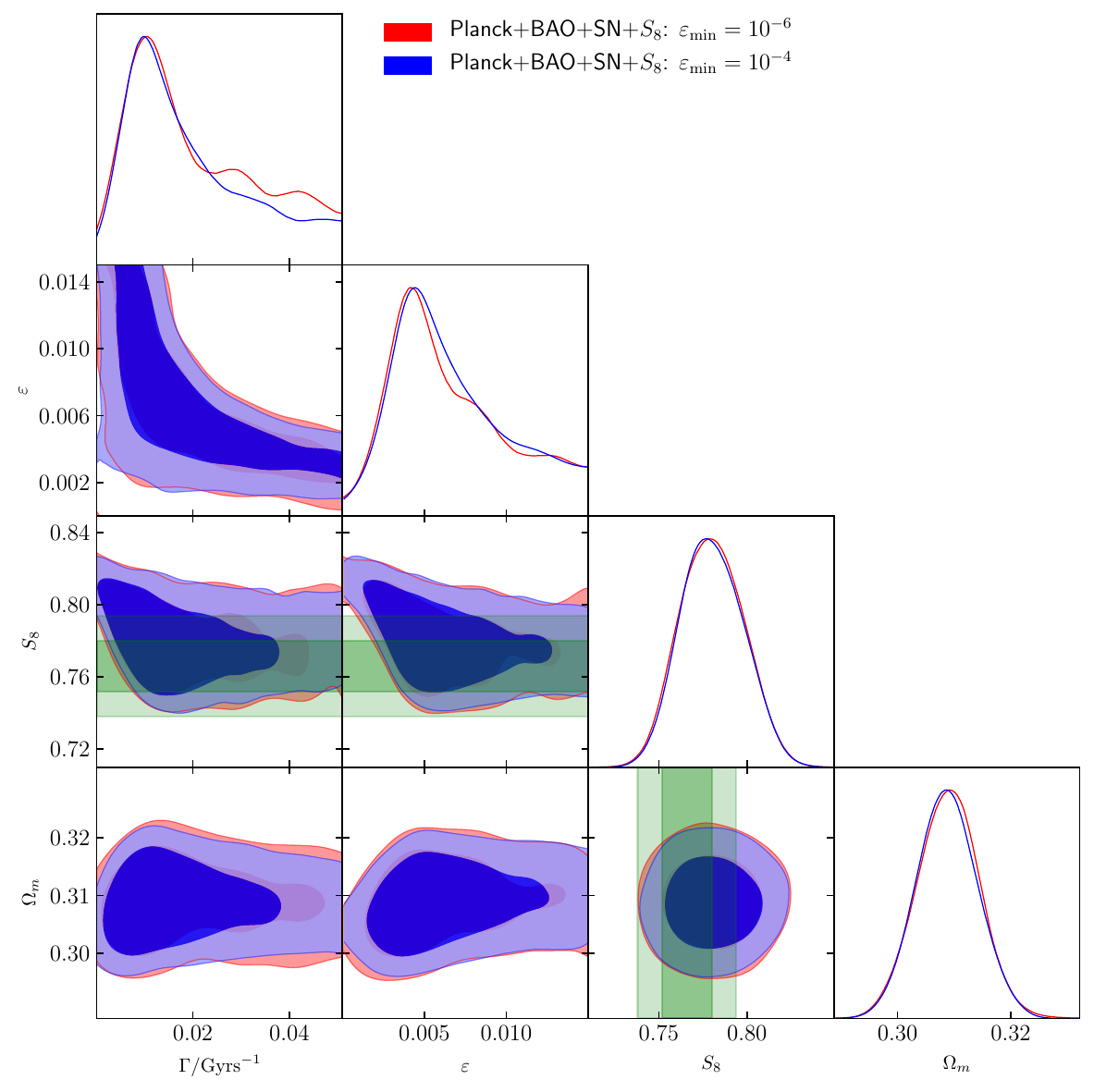}
        \includegraphics[width=1\columnwidth]{
        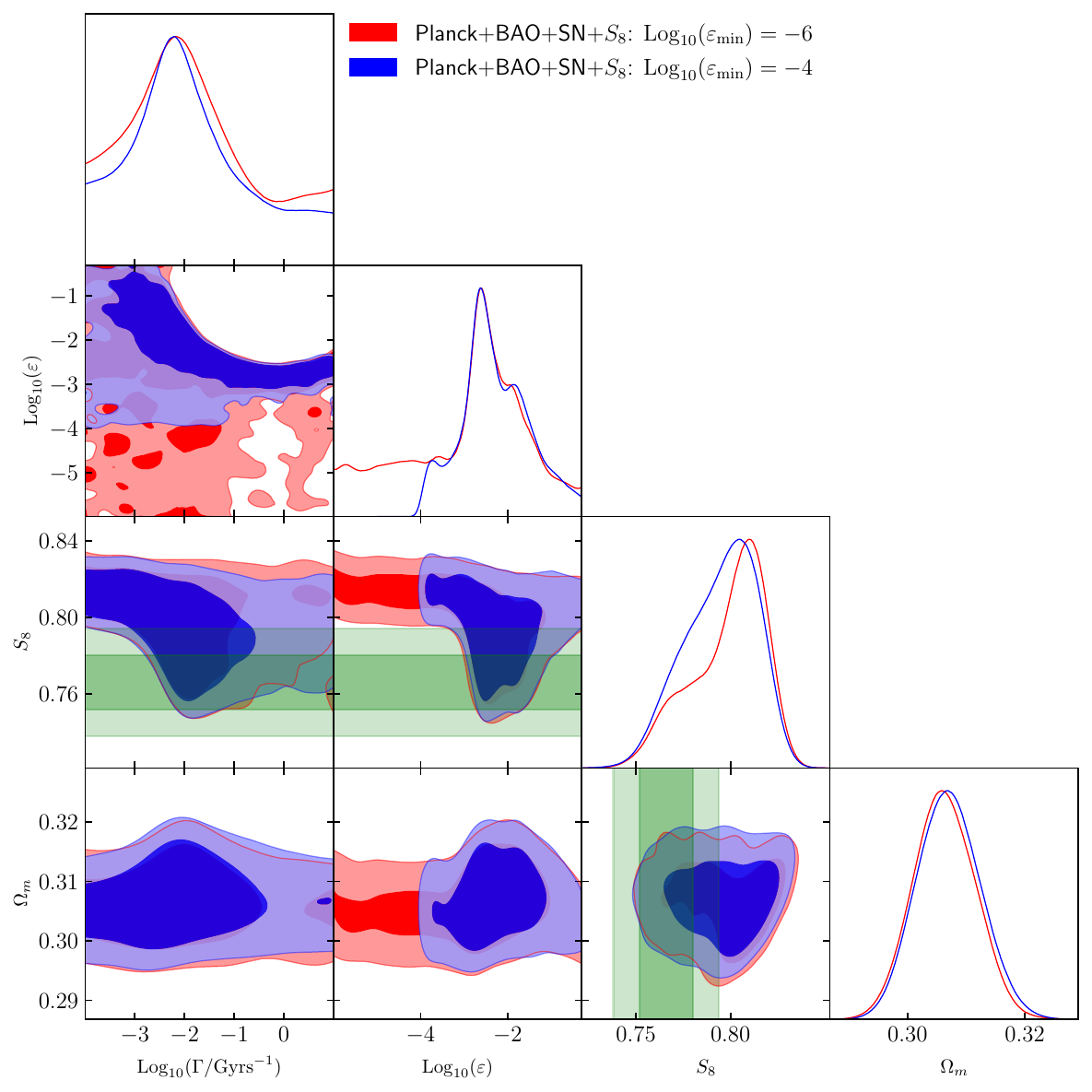}
    \caption{A comparison of the reconstructed 1D and 2D posteriors on $\{\Gamma,\varepsilon,S_8,\Omega_m\}$ when varying the lower bound on $\varepsilon$. We show the linear prior case on the left, and the log prior case on the right.}
    \label{fig:priors}
\end{figure*}

To check the impact of the lower bound of the prior on $\varepsilon$, we  perform additional runs with both the linear and the log prior changing the lower boundary from $\varepsilon=10^{-4}$  to $\varepsilon = 10^{-6}$.
We present the results in Fig.~\ref{fig:priors}. In the case of the linear prior, one can see that the results are unaffected by our choice of lower limit. 
In the log-prior case, the 2D-posterior of $\varepsilon-\Gamma$ is mildly affected by the lower bound (despite the somewhat `ugly looking' contours, we reach convergence of $R-1 < 0.01$), but the 1D-posteriors are largely unaffected. The main impact of extending the lower limit is that it gives more weight to the $\Lambda$CDM like parameter space, with a longer tail in the direction of low $\varepsilon$. Additionally, it leads to a somewhat sharper bi-modal $S_8$ 1D-distribution, rather than the heavy low-$S_8$ tail that we see with the prior $\log_{10}(\varepsilon)>-4$. Note that our tension metric is unaffected by the choice of lower-limit, given that it relies only on the best-fit parameters.
 
\end{document}